# Dynamic Priority Queue: An SDRAM Arbiter With Bounded Access Latencies for Tight WCET Calculation

Hardik Shah[1], Andreas Raabe[1] and Alois Knoll[2]

[1]ForTISS GmbH, Guerickestr. 25, 80805 Munich
[2]Department of Informatics VI, Technical University Munich, 85748 Garching

## Abstract

This report introduces a shared resource arbitration scheme "DPQ - Dynamic Priority Queue" which provides bandwidth guarantees and low worst case latency to each master in an MPSoC. Being a non-trivial candidate for timing analysis, SDRAM has been chosen as a showcase, but the approach is valid for any shared resource arbitration.

Due to its significant cost, data rate and physical size advantages, SDRAM is a potential candidate for cost sensitive, safety critical and space conserving systems. The variable access latency is a major drawback of SDRAM that induces largely over estimated Worst Case Execution Time (WCET) bounds of applications. In this report we present the DPQ together with an algorithm to predict the shared SDRAM's worst case latencies. We use the approach to calculate WCET bounds of six hardware tasks executing on an Altera Cyclone III FPGA with shared DDR2 memory. The results show that the DPQ is a fair arbitration scheme and produces low WCET bounds.

## 1 Introduction

There is a growing interest in using multi-core architectures in safety critical or mixed critical applications due to their superior performance per energy ratio. In these applications tasks have to satisfy stringent timing requirements imposed by deadlines. Failing to achieve a deadline may incur loss of lives or significant damage to property. Therefore, such applications must be certified to achieve all timing requirements before their industrial integration (e.g. Avionics). The guarantee of always achieving deadlines can only be achieved by timing verification through WCET analysis. For example, airbag, break-by-wire and stear-by-wire are Hard Real Time (HRT) applications of future electric cars with distinct bandwidth and latency requirements. When these applications are mapped to a multi-core processor, each application must achieve its deadline even in the presence of faulty co-existing applications. Hence, for WCET analysis worst behavior from the co-existing applications must be assumed.

In multi-core architectures, resources are often shared to reduce cost and exchange information. An off-chip memory is one of the most common shared resources. SDRAM is a popular off-chip memory currently used in cost sensitive and performance demanding applications due to its low price, high data rate and large storage. However, in time critical applications, SDRAM is rarely utilized due to its variable access latency. An asynchronous refresh operation and a dependence on the previous access make SDRAM access latency vary by an order of magnitude. Although it has cost and performance benefits, these access latency variations make SDRAM a difficult candidate for a precise WCET analysis. For a shared SDRAM, access scheduling latencies must be considered conservatively which decreases the tightness of the WCET bound.

The main contribution of this report is a novel arbitration scheme "Dynamic Priority Queue - DPQ" which satisfies distinct bandwidth requirements of masters with predictable worst case access latencies. The DPQ arranges masters in a priority queue depending on their access intensity to be fair as well as to provide low upper bound on latency to each master. An algorithm for WCET analysis of applications using the DPQ for resource sharing is provided as an API for an easy integration into third party WCET analyzers. The DPQ scheme is verified on an Altera Cyclone III FPGA using Micron DDR2-667 as a shared memory and the WCET bounds produced by the algorithm are found to be tight. Although, our test architecture is free from timing anomalies [15, 10], an outline of WCET analysis in their presence is provided in Sec. 5.1. For a quick industrial integration, Commercial Of The Shelf components are used with supporting circuitry for an FPGA implementation.

In the remainder of this report Sec. 2 discusses related work, Sec. 3 provides fundamentals, and Sec. 4 and Sec. 5 introduce our approach towards arbitration and latency analysis. Sec. 6 presents the evaluation and Sec. 8 concludes.

## 2 Related Work

Since the contribution of this report extends to fair arbitration as well as shared SDRAM latency analysis, we discuss the previous work in both the fields separately.

### 2.1 Arbitration Related

There has been a number of approaches to provide fairness, high throughput and worst case latency bounds in the arbiter especially in the networks domain. Weighted Round Robin (WRR) [12] is a work conserving arbiter where masters are allocated a number of slots within a round robin cycle depending on their bandwidth requirements. If a master does not use its slot, the next active master in the round robin cycle is immediately assigned to increase the throughput. Because of work conservation, masters which produce bursty traffic benefit at the cost of masters who produce uniform traffic [20]. Hence, to provide low worst case latency to any master, it has to be assigned more slots in the round robin cycle which leads to over allocation [5].

Deficit Round Robin (DRR) [22] assigns different slot sizes to each master according to its bandwidth requirements and schedules them in a Round Robin (RR) fashion. The major difference from traditional RR is that if a master cannot use its slot or part of its slot in the current cycle, the remaining slot (deficit) is added into the next cycle. In the next cycle, the master can transfer up to an amount of data equal to the sum of its slot size and the deficit. Thus, the DRR tries to avoid the unfairness caused to uniform traffic generators in the WRR. However, this approach leads to very high latencies in the worst case. For example, one master stays idle for a long time and gains high deficit. Afterwards, it contentiously requests the shared resource. Since it has gained a high deficit, it will occupy the shared resource for a long time incurring very high latencies to other masters.

Stratified Round Robin (SRR) [18] groups masters with alike bandwidth requirements into one class. After grouping masters into various classes two step arbitration is applied: interclass and intraclass. The inter class scheduler schedules each class $F_k$ once in $2^k$ clock cycles. *Hence, the lesser the $k$, the often the class is scheduled.* The intra class scheduler uses WRR mechanism to select the next master within the class. Due to more uniform distribution of bandwidth, SRR reduces the worst case latencies compared to the WRR. However, to achieve low worst case latency for a class $F_k$, $k$ must be minimized which leads to over allocation.

Priority Division (PD) [20] combines TDMA and static priorities to achieve guarantees and high resource utilization. Instead of fixing TDMA slots statically, PD fixes priorities of each master within the slot statically such that each master has at least one slot where it has the highest priority. Thus, masters have guarantees equal to TDMA and unused slots are arbitrated based on static priority to increase the resource utilization. This approach provides benefit over RR or WRR only if the response time of the shared resource is fixed. In the case of variable response time (e.g. SDRAM), this approach produces high worst case latencies.

Stein et al [24] introduces the Priority Based Budget Scheduler (PBS). In PBS, masters are assigned fixed budgets of access in a unit time (Replenishment Period). Moreover, master's are also assigned fixed priorities to resolve conflicts. Budget relates to master's bandwidth requirements while priority relates to master's latency requirements. Thus, coupling between latency and bandwidth is removed. The shared resource is granted to the active master with the highest priority which still has a budget left. At the beginning of a replenishment period, each master gets its original budget back. To avoid starvation of low priority masters, generally, masters with low budget are assigned high priority and masters with high budget are assigned low priority. However, due to the presence of priorities, PBS is fair to high priority masters and unfair to low priority masters. When all masters are executing HRTs (as outlined in the introduction), PBS results in large WCETs for low priority masters. Being the closest related work, in this report we will compare the PBS to our approach.

Akesson et al [5] introduce a Credit Controlled Static Priority (CCSP) arbiter. The CCSP also uses priorities and budgets within the replenishment period. But, instead of using frame based replenishment periods, masters are replenished incrementally for fine grade bandwidth assignment. However, due to the presence of the priorities, large WCET bounds for lower priority masters are produced.

## 2.2  SDRAM Related

PRET [14] is a cycle accurate repeatable time machine. The PRET machine runs four hardware threads simultaneously in an interleaved manner. To predict SDRAM latencies, the PRET DRAM controller [19] is used. The PRET DRAM controller views banks and ranks of an SDRAM as individual memory resources. Each thread has exclusive access to its private resource in a time triggered manner. Since resources are scheduled in a Round Robin (RR) fashion, each resource gets sufficient time to satisfy SDRAM timing requirements until it is accessed again. The controller achieves predictability by sharing on-chip memory resources and keeping off-chip memory resources private. Due to the tremendous cost of on-chip resources, this approach is applicable only where masters share only a tiny fraction of data. In state-of-the-art technology, sharing even medium sized data (e.g. a single VGA frame) it will be unfeasibly expensive. The backend of the PRET DRAM controller uses bank interleaving (BI) together with Rank Interleaving (RI) to hide the access latencies. RI together with BI reduces Read/Write switching penalties. Currently, our implementation utilizes only one rank. Therefore, we are using only BI (Sec. 3.4). This is not a restriction of our approach. RI can simply be achieved by using an additional rank and doubling address space and access granularity.

PREDATOR [3] applies a holistic approach investigating the entire spectrum of system design (architecture, compiler, operating system, software development, etc.) to come up with a predictable and efficient system solution. However, SDRAM latency analysis is not included. MERASA [26] targets to build a predictable and efficient multi-core architecture for mixed critical applications. The SDRAM is accessed through an Analyzable Memory Controller (AMC) [17]. The AMC uses the RR policy and applies BI to access the shared SDRAM. By theoretical analysis, latency parameters are extracted for the WCET calculation. The RR policy with one transfer per master cannot satisfy the need of different bandwidth requirements. Weighted Round Robin [12] satisfies this need but, as discussed before, to achieve low latency, high bandwidth must be assigned to the according master, leading to over allocation.

Heithecker et al [11] propose BI with two level arbitration using RR scheme to bound the maximum latency of requesters. Requesters are assigned high and standard priorities. High priority requesters are scheduled out of order in the RR queue. After each high priority request, one standard priority request is scheduled to avoid starvation. The approach excludes SDRAM latency analysis and inherits drawbacks of RR scheme as discussed above.

Bhat et al [7] propose a periodic software task that refreshes the SDRAM and thus, removes asynchronism related to hardware controlled refresh. When such approach is applied to multi-core architectures, only the core running the *refresh task* will benefit. Other cores still have to consider it as an asynchronous refresh operation. Bourgade et al [8] propose a static analysis of the SDRAM row buffer in the open page policy (cf. Sec. 3.3). The analysis tracks worst case cache misses on an execution path and predicts whether the cache miss incurs SDRAM row buffer hit or miss. This approach is valid only for unshared SDRAMs. In modern systems, SDRAM is shared between i-cache, d-cache, Direct Memory Access (DMA) engines etc. Moreover, refresh penalty analysis is not included.

Akesson et al [4] propose a predictable SDRAM controller using BI and CCSP [5]. High priority is assigned to masters executing latency sensitive applications while masters requiring high bandwidth are assigned higher budget. Thus, the coupling between latency and bandwidth is removed. However, worst case SDRAM latency analysis for individual access is not provided.

Shah et al [21] calculate the WCET of an application using worst case memory traces. The authors provide in-depth latency analysis of BI and implement a PBS [24] on an Altera FPGA. By applying aggressive alternating accesses to the DDR2 memory, worst case latency parameters are extracted and used later for WCET calculation. Their results show the unfairness to the lower priority masters when PBS is applied for sharing an SDRAM. In this report we will use their results for comparison with the DPQ. Staschulat et al [23] provide data flow models for shared memory access latency analysis for PBS arbitration. However, in their approach there is only one high priority task and all other tasks are of low priority. Hence, their approach cannot be used for a mixed critical system with multiple HRT applications.

Apart from the above mentioned related work, Kelter et al [13] and Dasari et al [9] present approach to calculate WCET of applications in multi-core architectures using shared memory. Kelter et al [13] assume that L2 cache is shared among cores using TDMA arbitration. The WCET is then calculated using the offset of the arrival time of memory requests. Dasari et al [9] presents an approach to calculate Worst Case Response Time (WCRT) of a task without knowing the underlying shared memory arbitration scheme. To achieve worst case time for completion of a single memory request, memory is accessed known number of times. Then the time consumed for these accesses is divided by the number of accesses. Thus, the average latency value is considered as the worst case latency for a single access. There are two drawbacks of this approach considering SDRAM as an off-chip memory. i) For an off-chip memory like SDRAM, the current access latency heavily depends on previous access [2, 21]. ii) Read and write have different latencies (cf. Sec. 3.5). Hence, reading the off-chip memory number of times and taking the average of latency is not the actual worst case. Moreover, this approach cannot deal with uncertainty caused by Refresh. Lv et al [16] analyzes local cache behavior using Abstract Interpretation (AI). Based on the analysis, timed automaton of the program is constructed to obtain precise timing information of cache miss. The shared bus is also modeled using timed automata. The timed automaton models of the bus and programs running on separate cores are explored using the UPPAAL model checker to find the WCETs of the respective programs. This approach works only if all the information about the co-existing applications are available. Hence, change in one application (e.g. a bug fix) forces re-analysis of all the co-existing applications. Moreover, they assume an ideal memory controller which always responds with the constant latency. In this report we analyze applications in isolation. Moreover, we use non-ideal, COTS DDR2 memory controller.

## 3 Background

### 3.1 Application Execution

Fig. 1a abstracts behavior of an application execution assuming in-order execution. At first, application code/data is read or written. If the accessed data is on-chip (e.g. cache, registers etc) it is processed immediately. Otherwise, data is fetched from the off-chip memory (here, SDRAM) and then processed. Off-chip memory is accessed transparently through a cache. The cache is organized in the form of cache lines as depicted in Fig. 1b. When a cache miss/write-back occurs, an entire cache line (here, 32 bytes) is read from or written to the SDRAM. The SDRAM is accessed only if a cache miss or cache write-back occurs. During on-chip operations the SDRAM stays idle. In the following we will denote these idle times as `OnChipProcTime` $\in Z^+$.

Note that in [4, 5] and in many other literature, network/real-time calculus [25] is used for analysis. This analysis assumes that the requests are independent of each other. However, in case of execution on an in-order core, the next request is not issued until the previous request is served. Neglecting this inter dependence will lead to a largely over estimated WCET bounds. Moreover, real-time calculus does not differentiate between *read* and *write* requests. Our approach analyzes read and write requests individually for precise latency calculation. Hence, we have omitted the use of network/real-time calculus.

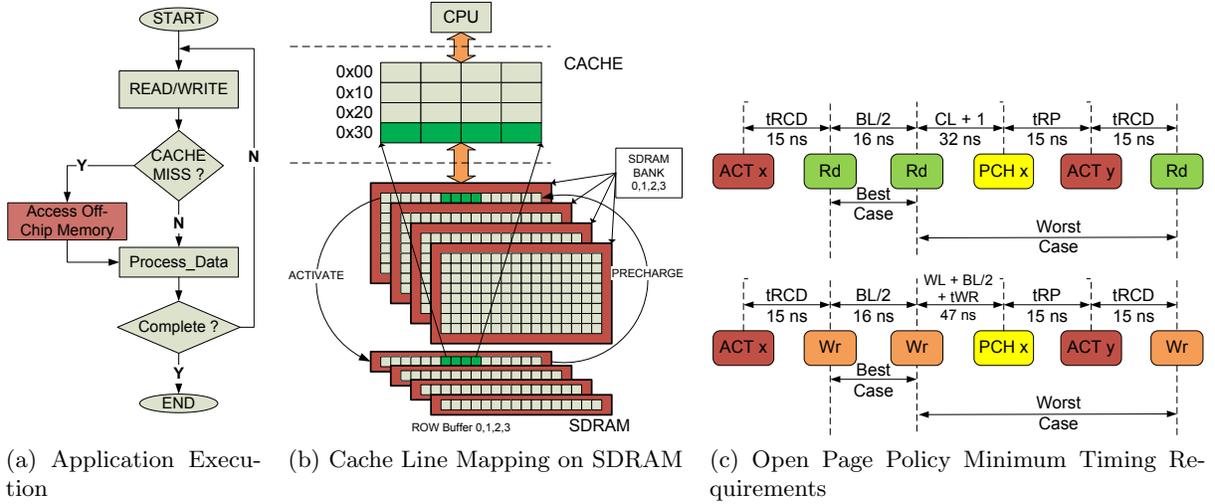

(a) Application Execution  (b) Cache Line Mapping on SDRAM  (c) Open Page Policy Minimum Timing Requirements

Figure 1

## 3.2 SDRAM Fundamentals

SDRAMs have a bank architecture where each bank stores data in a 2D array of rows and columns as depicted in Fig. 1b. Cache lines are mapped to the rows of banks. Each row contains multiple cache lines. Before reading or writing a cache line, the entire row containing that cache line is copied into the row buffer. This operation is called *Row Activation* (ACT). Data is now read from or written to the row buffer in a burst fashion with low latency. When data from another row is requested, the content of the buffer is copied back to the bank. This operation is called Precharge (PCH). Subsequently, the new row is activated. At compile time it is unknown which row will be in the row buffer at a certain point of the execution. Being conservative, an absence of the currently accessed row from the buffer is assumed for worst case latency analysis. This assumption incurs large WCET bounds (cf. Sec. 3.3) of the application. Moreover, SDRAMs must be refreshed periodically. During a refresh operation the SDRAM cannot be accessed. Since the refresh operation is controlled by hardware, its occurrence time is unknown to the application. This uncertainty further degrades WCET bounds.

## 3.3 Open Page Policy

Fig. 1c depicts various timing parameters that must be satisfied before applying commands to the SDRAM. After row X has been activated, read or write commands can be applied with minimum latency. But if row Y of the same bank has to be accessed, at first row X has to be precharged and then row Y can be activated.

The values of the parameters shown in the figure are for DDR2-667 memory. Here, BL (Burst Length) = 4. At 125 MHz, clock period is 8 ns and hence, BL/2 clock period = 16 ns. CL + 1 = 4 clocks = 32 ns. WL = CL - 1 = 2 clocks = 16 ns. tWR = 15 ns. Thus, using an open page policy and assuming worst case always, will result in a large WCET bounds. Note, that the figure shows worst case when the SDRAM is unshared. For a shared SDRAM, the worst case latencies can be

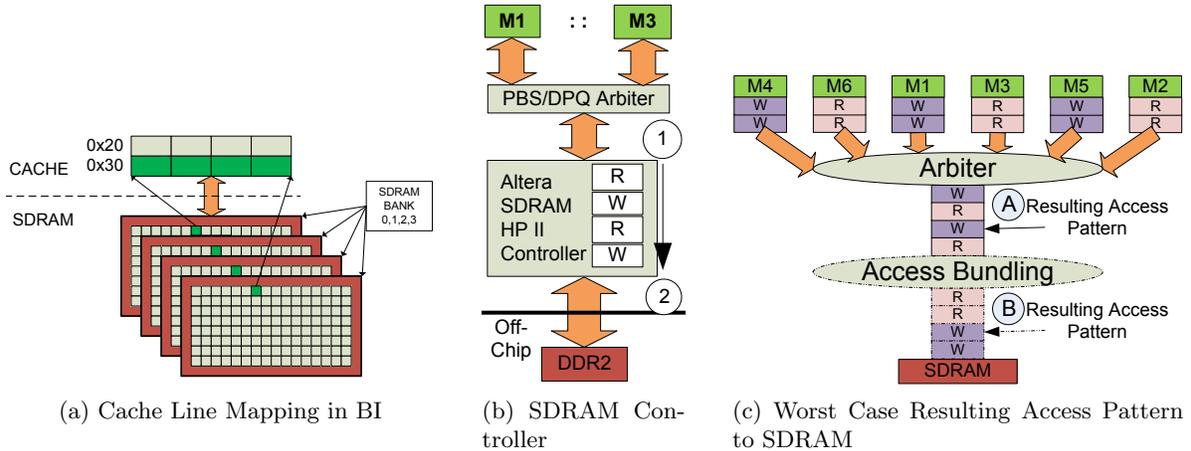

Figure 2

(a) Cache Line Mapping in BI

(b) SDRAM Controller

(c) Worst Case Resulting Access Pattern to SDRAM

as many as n (n = Number of Masters) times larger. In such case difference between best case and worst case for a read access will be ≈ $4n$ times and for a write access ≈ $5n$ times. Latencies associated with refresh and alternating accesses further increase the worst case latencies (not shown in the figure).

### 3.4 Bank Interleaving - BI

To mitigate the penalty associated with row buffering, BI can be applied. BI splits all off-chip accesses into small chunks such that each chunk is executed on a unique bank and each bank of the device is assigned a unique chunk. The bank is then precharged at the earliest possible time. While data is provided from one bank, another bank can be activated in the background. The resulting cache mapping to the SDRAM is depicted in Fig. 2a. The alternating read/write accesses produce the worst case latencies when applying BI [21]. Thus, when analyzing a read (write) access we will assume it is interfered by a write (read). For more details of latencies associated with SDRAM and BI, we refer readers to [21].

Note, that although modern compilers can optimize the access sequence of the core and group read accesses and write accesses together, in multi-core architectures, the resulting pattern to the SDRAM may still be alternating as shown in Fig. 2c at point **A**. Here, each master generates either read or write access sequence. However, due to the arbitration, the resulting accesses to the SDRAM can be of alternating type. To increase the average case performance access bundling can be applied where accesses are re-ordered to minimize Read/Write switching as shown in Fig. 2c at point **B**. However, such an access re-ordering increases worst case latency bounds. For example, if one of the cores wants to do a write access when all other cores are intensively doing read accesses, the write access has to wait for a longer time. Hence, we have not implemented access bundling techniques.

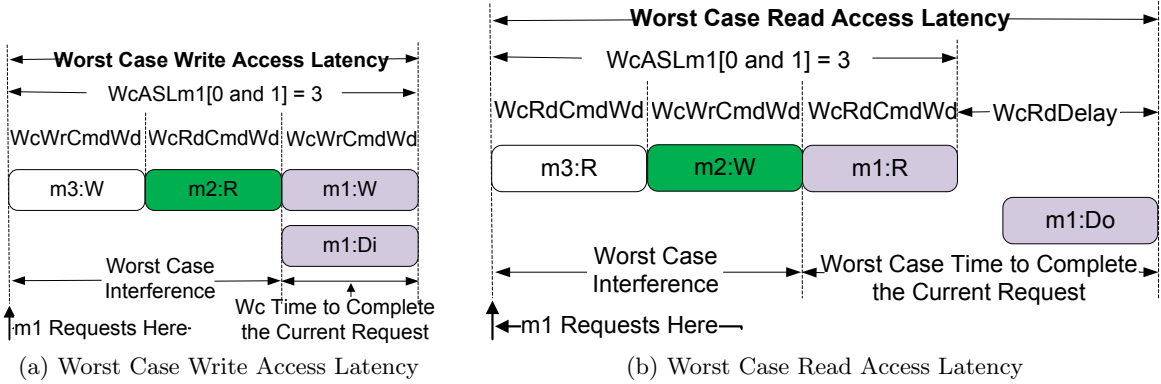

Figure 3: Read and Write Access Parameters

## 3.5 Read/Write Access Parameters

Fig. 3 elaborates read and write accesses of master m1 and interference experienced by them. On a read access (Fig. 3b), m1 puts address and read command (m1:R) on the command bus. The worst case width of such a read command is denoted by $W_cRdCmdWd$. After some delay the memory puts the requested data onto the data bus. The worst case time from completion of read command to the reception of the entire data is denoted by Worst Case Read Delay ($W_cRdDelay$). On a write access (Fig. 3a) m1 puts address, write command (m1:W) and write data altogether at the same time on their respected buses. The worst case width of a write command is denoted by $W_cWrCmdWd$.

## 4 Dynamic Priority Queue - DPQ

In DPQ, masters are assigned a fixed budget per replenishment period at design time. The fixed budget relates to the master's bandwidth requirements. For example, a master processing video frames will be assigned more budget than a master processing audio frames since the former needs more bandwidth than the latter. In DPQ, the priority of a master, at any time, depends on its position in the queue. The DPQ arbiter always starts searching for a master from the left of the queue as depicted in Fig. 4. The arbiter selects the first requesting master in the queue which has a budget left. The selected master is granted the shared resource, its budget is reduced by one and it is transfered to the end of the queue. Masters which are on the left of the granted master remain as they are and masters which are on the right of the granted master are shifted one step to the left. In other words, masters which had a higher priority than the granted master keep their priorities unchanged. Masters which had a lower priority than the granted master increase their priority to the next higher level. The recently granted master receives the lowest priority.

In Fig. 4, at point **A**, though master m3 is requesting and it has the highest priority (first in the queue), it is not granted since it is ineligible in the current replenishment period (BudgetLeft = 0). Here, master m2 is granted, its budget is reduced by one and it is enqueued at the end. This operation shifts master m1 to the next higher priority level. Note, that the status of m3 remains unchanged. At point **B**, m1 being the only eligible and requesting master, it is granted and enqueued at the end after reducing its budget by one. At point **C**, a new replenishment period starts and all masters are re-assigned their initial budget making all of them eligible. The remaining budget of the previous replenishment period is discarded. At point **D**, `m3` is granted since it has the highest priority. Thus, when some master is not granted for some time due to ineligibility or inactivity, it will automatically reach a higher priority level (provided that the other masters are active during that time).

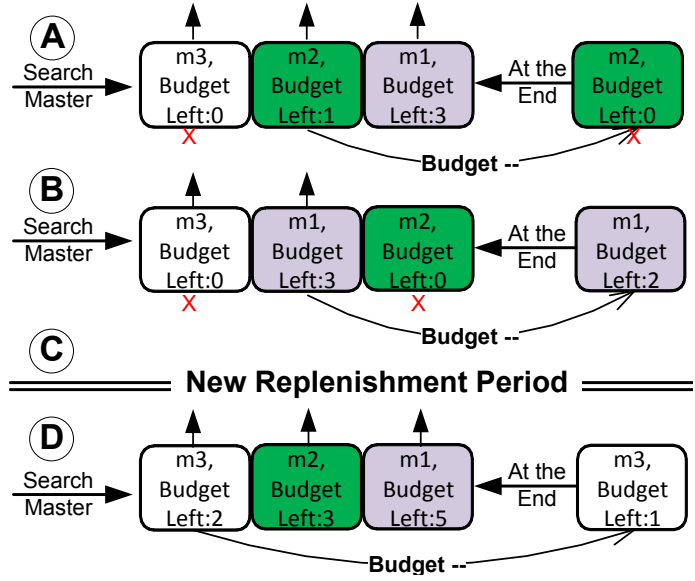

Figure 4: Dynamic Priority Queue. n = 3, MaxBudget = 5 Budget[m1,m2,m3]={5,3,2}.

**Side Note:** One advantage of the DPQ scheme is that if the access patterns of other masters in the system are known, worst case access latency prediction of the master under investigation can be greatly improved. By accessing, other masters actually help the master under investigation to gain higher priority levels. For example, consider a mixed critical system with 25 frames per second video application. The application will be either turned off or on. If it is turned off it is not causing any interference and if it is turned on, it does regular accesses to maintain its data rate and help the master under investigation to achieve a higher priority level. In both the cases, the access latency of the master under investigation will improve. However, in this report we analyze applications in isolation assuming the worst behavior of other masters.

Let, `n = NumberOfMasters`.

$$R_p = \left\lceil \frac{(W_c RdCmdWd + W_c WrCmdWd)}{2} \right\rceil \times \sum_{i=1}^{n} Budget[i] \qquad (1)$$

The replenishment period is dimensioned such that theoretically all masters could exploit their entire budget. Equation (1) gives the size of a replenishment period in clock cycles considering an alternating read/write accesses. It should be noted that in practice, read accesses significantly outnumber write accesses during application execution. However, for any cache miss (read or write), at first, the evicted cache line (dirty line) is written back to the main memory and new cache line is read from the main memory. Thus, cache miss produces main memory write and main memory read as long as the evicted cache line is dirty. Moreover, when analyzing in isolation, the worst case must be assumed for estimating the absolute WCET. In the case of probabilistic WCET analysis [6] less conservative assumptions can be made. However, probabilistic WCET analysis is out of the scope of this report.

## 4.1 Size of Replenishment Period

In equation (1) the equality must hold for fairness and high resource utilization. For example, size of the replenishment period is $R'_p < R_p$. Consider that all masters generate accesses in such a way that the resulting accesses to the shared SDRAM is of alternating type (Fig. 2c). Moreover, assume that all the masters have `OnChipProcTime = 0`. Here, at least one master will get less share than its allocated budget in a replenishment period due to the condition in equation (2).

$$R'_p < \left\lceil \frac{(W_c RdCmdWd + W_c WrCmdWd)}{2} \right\rceil \times \sum_{i=1}^{n} Budget[i] \qquad (2)$$

If $R'_p > R_p$ than the SDRAM will stay idle for at least $(R'_p - R_p)$ clock cycles in each replenishment period. Thus, bandwidth utilization will be degraded. Moreover, latency for each master also increases since it has to wait longer to be replenished.

## 4.2 Offline Budget Assignment

In the DPQ, PBS and CCSP budgets are defined at design time and do not change at run time. Hence, in the case of multi threaded execution where each thread has different bandwidth requirements, threads with similar bandwidth requirements must be grouped and executed on the same core. On one hand this is a clear drawback and dynamic budget assignment should be applied. On the other hand dynamic budget assignment dynamically changes the size of replenishment period (equation (1)) and hence the worst case latency bounds. Thus, the WCET produced using such latency bounds is very high.

## 4.3 Worst Case Interference

Fig. 2b depicts the architecture and Fig. 4 illustrates DPQ operation with three masters. For calculating the worst case interference, each master must consider the worst behavior of other masters in the system. For example, in Fig. 4 at point **D** all masters are eligible. The position of any master in the queue is unknown at analysis time and hence the master under-investigation (here, `m1`) must be assumed to be at the end of the queue. Thus, the maximum possible number of interfering accesses for the first access of `m1` are (n - 1). If `m1` requests a second access in the same replenishment period, again it must be assumed to be at the end of the queue with a maximum possible number of interfering accesses (n - 1). For the third access of `m1` in the same replenishment period, only one interfering access from `m2` has to be considered since interference from `m3` has already been considered twice. Hence, `m3` is ineligible in the current replenishment period (point **A** in the Fig. 4). For the forth and fifth requests of `m1` in the same replenishment period, interference from other masters is zero since all other masters are ineligible (point **B** in the Fig. 4). We denote these worst case interfering accesses for master m as $I_m$ and calculate its value as an array of length $MaxBudget = 5$. $I_{m1}[5] = (2, 2, 1, 0, 0)$, $I_{m2}[5] = (2, 2, 1, X, X)$ and $I_{m3}[5] = (2, 2, X, X, X)$. Note that master `m2` and `m3` have budgets of 3 and 2 respectively. Hence, their $I_m$ array contains $'X'$s which denote ineligibility of the respective master in the replenishment period (Algorithm 1).

**Algorithm 1** Worst Case Access Sequence Length - DPQ
─────────────────────────────────────────────
1: $OriginalBudget[n] = \{...\}$
2: **repeat**
3:    $Temp[] = OriginalBudget[]$
4:    **repeat**
5:      **if** $Temp[j] = 0$ **then**
6:        $W_cASL[j][i] = X$             # ineligible in the current replenishment period
7:      **else**
8:        $acc = 0$
9:        **repeat**
10:          **if** $Temp[k] > 0$ **then**
11:            $acc = acc + 1$
12:            $Temp[k] = Temp[k] - 1$
13:          **end if**
14:        **until** $k \leq n$
15:        $W_cASL[j][i] = acc$
16:      **end if**
17:    **until** $i \leq MaxBudget$
18: **until** $j \leq n$
─────────────────────────────────────────────

During execution, it may be possible that instead of the first and the second accesses of `m1` in a replenishment period, other accesses are interfered by (n - 1) accesses. In the DPQ, an earlier access request cannot produce longer execution time than had it been requested later. Thus, the DPQ does not introduce any timing anomaly and the architecture is assumed to be without timing anomalies. Hence, commutativity holds and the worst case interference can be assumed in the beginning of a replenishment period. In any case, $I_m$ must satisfy conditions of equations (3) and (4).

$$\sum_{i=1}^{Budget[m]} I_m[i] \leq \sum_{j=1}^{n} Budget[j] \qquad , j \neq m, \qquad (3)$$

$$I_m[i] \leq (n-1) \qquad , \forall i \in Budget[m] \qquad (4)$$

### 4.4 Worst Case Access Latency

Fig. 3 depicts worst case access latencies for read and write accesses. To compute worst case latency, it must be assumed that the interfering accesses also take the worst case time to complete. In other words, the worst case latency is the total worst case time to complete the interfering accesses plus the current access. For worst case latency calculation, we consider the current access to be a sequence of accesses formed by interfering accesses and the current access. The length of such a sequence is denoted by Worst Case Access Sequence Length ($W_cASL$) for $x^{th}$ access of

**Algorithm 2** Worst Case Access Latency for master **M**

1: /*PreTime = Previous Access Time*/
2: /*Rp = Replenishment Period*/
3: /*NewRp = New Rp*/
4: /*CrtPos = Current Position in Rp*/
5: /*RemRp = Remaining Rp*/
6: /*tRef = Time to Refresh*/

7: $CalcLat(CurrentTime, AccessType, Master = M)$
8: {
9: $RemRp = 0$
10: $OnChipProcTime = CurrentTime - PreTime$
11: **if** $used = Budget[M]$ **then**
12:    $used = 0$
13:    $NewRp = true$
14:    **if** $Rp > (CrtPos + OnChipProcTime)$ **then**
15:      $RemRp = Rp - (CrtPos + OnChipProcTime)$
16:      $CrtPos = 0$
17:    **else**
18:      $CrtPos = (CrtPos + OnChipProcTime) - Rp$
19:    **end if**
20: **else**
21:    **if** $(CrtPos + OnChipProcTime) \geq Rp$ **then**
22:      $used = 0$
23:      $CrtPos = (CrtPos + OnChipProcTime) - Rp$
24:      $NewRp = true$
25:    **end if**
26: **end if**
27: **if** $AccessType = Write$ **then**
28:    $lat = WrTime(W_cASL[M][used])$
29: **else**
30:    $lat = RdTime(W_cASL[M][used])$
31: **end if**
32: $used = used + 1$
33: **if** $NewRp$ **then**
34:    $CrtPos = CrtPos + lat$
35: **else**
36:    $CrtPos = CrtPos + OnChipProcTime + lat$
37: **end if**
38: $lat = lat + RemRp$
39: $tRef = tRef + lat + OnChipProcTime$
40: **if** $tRef \geq tREFI$ **then**
41:    $tRef = tRef - tREFI$
42:    $lat = lat + tRFC$
43: **end if**
44: $PreTime = lat + CurrentTime$
45: $return(lat)$
46: }

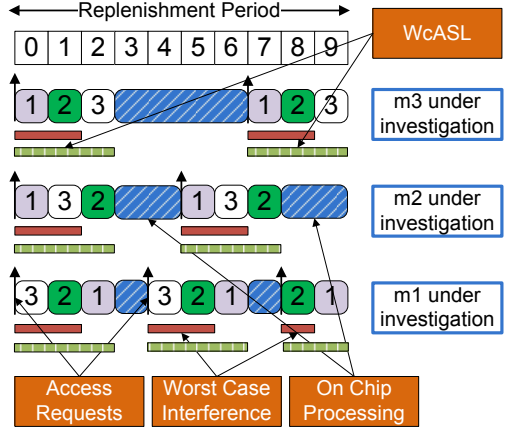

Figure 5: Analysis View of a Replenishment Period in DPQ

master $m$, $W_cASL[m][x] = I_m[x] + 1$. We also assume that $W_cASL[m][x]$ is formed by alternating read-write accesses. Algorithm 1 calculates $W_cASL[j][i]$ for the $i^{th}$ request of master $j$ in a single replenishment period.

After calculating $W_cASL[j][i]$ for each master, tagging of each request as a first, second etc in the current replenishment period is required. Algorithm 2 tags each access and returns the worst case latency. At first, `OnChipProcTime` is calculated by subtracting the completion time of the previous request from the current simulation time. If the master is ineligible then it has to wait till the next replenishment period. Hence, the current access will be moved to the next replenishment period where it will be considered to be first access of M in replenishment period. The remaining time of the current replenishment period will be added to the latency. If the master is eligible, but due to a long `OnChipProcTime`, has crossed the replenishment period boundary $((CrtPos + OnChipProcTime) \geq Rp)$ then also the current access will be considered as a first access in the new replenishment period. Functions $WrTime$ and $RdTime$ calculate the worst case latency of the current access by using parameters of Fig. 3. The latency and `OnChipProcTime` are added to the refresh counter $(tRef)$. If the refresh counter exceeds $tREFI$, the penalty of a refresh $(tRFC)$ is added to the latency.

For BCET calculation, we assume that the master is scheduled immediately and all refresh are masked by an `OnChipProcTime`. Hence, no refresh penalty is considered.

### 4.5 Analysis View of a Replenishment Period

Fig. 5 depicts a single replenishment period with three masters `m1`, `m2` and `m3`. Budget[m1,m2,m3] = [5,3,2]. From equation (1), replenishment period must be long enough to allow 10 alternating accesses. As shown in Fig. 5, for the worst case latency analysis, the master under investigation is always assumed to be at the end of the queue when it requests. Moreover, it is assumed that when the master under investigation requests, all other masters also request and hence the master under investigation experiences the maximum interference. If the `OnChipProcTime` of master under investigation is zero, then it will be able to exploit its entire budget in a replenishment period even under the above mentioned conservative assumptions. By using a Scratch Pad Memory (SPM) and a Direct Memory Access (DMA) zero `OnChipProcTime` can be achieved. This will lower the WCET bound of the application. The `OnChipProcTime` is excluded from the definition of replenishment period to provide high resource utilization.

The conservative assumptions mentioned above, will consider the late accesses in a replenishment period as the early accesses in the next replenishment period during analysis. For example, in Fig. 5, `m2` will not be able to do its third access in the replenishment period. The third access of `m2` crosses the replenishment period boundary. Hence, the condition $(CrtPos + OnChipProcTime) \geq Rp$ of Algorithm 2 is true. This access will be considered as the first access in the next replenishment period where it will experience a high interference (cf. Sec. 4.3). The same is true for the forth and the fifth accesses of `m1` in the figure. The master which does more accesses is more probable to do accesses late in the replenishment period, considering uniform distribution. Hence, it is more penalized by the conservative assumptions mentioned above.

During execution, most of the times, such highest interference does not occur. The `OnChipProcTime` shown in the Fig. 5 are overlapped by access requests from other masters. Hence, masters are able to exploit their entire budget in a replenishment period. But, such favorable scenario cannot be guaranteed when analyzing under isolation and thus, conservative assumption of the highest interference must be considered for the worst case latency analysis.

## 4.6 Pessimism

Algorithm 2 assumes that all the possible refreshes that may occur during an application execution, interfere with the accesses of that application. In reality, it is not always true. Many refreshes occur while the application is executing from a cache or registers. Such refreshes do not interfere and hence do not contribute to the execution time of the application. Thus, Algorithm 2 over estimates latencies w.r.t. the refresh. Only this pessimism contributes to the overestimation of WCET bounds in our approach. The assumption of interference from alternating accesses is not pessimistic, but conservative, since we assume no knowledge of access patterns of other masters. We only know their budgets in a replenishment period which are fixed at design time. Although we know everything about the refresh ($tREFI, tRFC$), interference caused by it is assumed pessimistically since the coincidence of a memory access and a refresh is practically impossible to predict. To estimate the worst case overestimation, we assume that the application is not interfered by refreshes at all ($tRFC = 0$). We denote the WCET calculated considering no refresh interference by $WCET^{nr}$. In the following, the ratio $WCET/WCET^{nr}$ will be considered to be the worst case over estimation (pessimism) of our approach.

## 5 Tooling

We implemented our SDRAM Timing Analyzer (Algorithm 1 and Algorithm 2) as an API for integration into third party WCET analyzers. The parameters depicted in Fig. 3 are directly measured on an FPGA using an on-chip logic analyzer (SignalTap II) and supplied to the API. As discussed before, our API can supply the best case latency as well as the worst case latency of a single access.

To facilitate state reduction in third party WCET analyzers, an additional API `MaxMissCnt()`

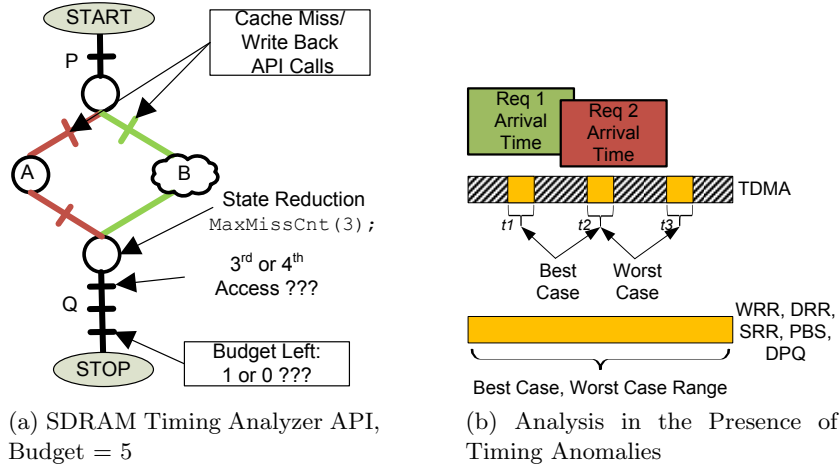

(a) SDRAM Timing Analyzer API, Budget = 5

(b) Analysis in the Presence of Timing Anomalies

Figure 6

and an internal access counter of `TotalAccesses` are included in the algorithm. As depicted in Fig. 6a, there are paths PAQ and PBQ to reach from `START` to `STOP`. The application is executed on `m1` which has a budget of 5. After both the paths converge, the access latencies of cache misses on branch Q depend on which path was taken. Let us assume that during path analysis, the WCET analyzer notices that the path PB has a longer execution time due to a big calculation block (2 cache miss + time for calculation) than path PA (3 cache misses). Hence, at the point of convergence, the WCET analyzer may apply state reduction to consider only path PB for further WCET calculation. Nevertheless, it may be possible that PAQ is longer than PBQ since the former has 6 cache misses. To serve 6 cache misses `m1` needs at least two replenishment periods while the path PBQ may finish in only one replenishment period leading to a false WCET bound. To mitigate this problem, at the point of convergence, the WCET analyzer has to call `MaxMissCnt()` with the maximum cache miss count (here, 3) from all the converging paths. The SDRAM timing analyzer then reduces `Budget[m1]` (cf. Algorithm2) by the difference of maximum miss count supplied by the WCET analyzer and the internally calculated `TotalAccesses` (2 in case of path PB). Since the early accesses in a replenishment period experience worse interference than the late accesses (cf. Sec. 4.3), the variable `used` of Algorithm 2 is not increased. The reduction of `Budget[m1]` is done only for the current replenishment period. Thus, in the case of state reduction, the cache misses in the Q branch will have access latencies according to path PB. The last access will be considered to be the $5^{th}$ access in the current replenishment period where `m1`'s budget is $5 - (3 - 2) = 4$. Hence, this access becomes the first access in the next replenishment period and latency will be calculated accordingly. Now, in both the cases from `START` to `STOP` at least two replenishment periods are required. This approach will enable trading off precision of the WCET bound for faster analysis.

## 5.1 Timing Anomalies

In the presence of timing anomalies, the WCET analyzer has to call the API considering the range of all possible $CurrentTime$ values. The API then returns $B_c$ and $W_c$ latency values for each value of $CurrentTime$. Then the WCET analyzer has to pick the combination of a $CurrentTime$ and a latency value which produces the longest over all execution time of the program. Thus, the computation of the WCET becomes exhaustive which is a typical case of the presence of timing anomalies.

Since DPQ, PBS, CCSP, WRR, SRR, DRR schedule the next master dynamically, the $[B_c, W_c]$ range is large. For example, consider the overlapping arrival time of two requests in the presence of timing anomalies as depicted in Fig. 6b. Here, for TDMA, in the best case two accesses will be scheduled in the first and second slots of the figure. In the worst case these two accesses will be scheduled in the second and third slots. Moreover, precise start time of these slots are known (*t1, t2, t3*). Hence, the third party WCET analyzer has to consider only these three values for two accesses depicted in the Fig. 6b. While for dynamic arbitration the entire range of $[B_c, W_c]$ has to be considered. Kelter et al [13] presents an approach for calculating WCET using TDMA offsets in the multi-core architecture. However, due to efficiency reasons, TDMA arbitration is rarely used [9].

## 6 Test Setup

In the following the performance of the DPQ and the PBS are compared. We connected six hardware traffic generators to an SDRAM controller running on an Altera Cyclone III FPGA, as depicted in Fig. 2b. The Altera High Performance Controller II (HP II) does not support BI in the native mode. Therefore, we implement BI using access splitting and user controlled auto precharge [1]. The HP II contains internal FIFOs for commands and data. These FIFOs affect access latencies depending on their status. In case of a *write* transfer, the master proceeds with its operation while command and data move from `point 1` to `point 2` in Fig. 2b and are subsequently written to the SDRAM. For read transfers, time required by a *read* command to reach from `point 1` to `point 2` must also be considered. After arriving at `point 2`, latencies of Fig. 3 must be considered.

We obtained safe values of worst case latency parameters ($W_cRdCmdWd$, $W_cWrCmdWd$ and $W_cRdDelay$) by generating continuous alternating traffic to the SDRAM and measuring the parameters at `point 1` in Fig. 2b using an on-chip logic analyzer. Since `point 1` is the only visible point to the user, it is safe to extract all latency parameters from there. Using Signaltap II, we also found that the Altera SDRAM controller issues a refresh every $tREFI \pm 20$ clock cycles depending upon ongoing transfers. To avoid this variability, we implemented a user controlled refresh circuit [1] that closes all channels at `point 1` slightly before tREFI to empty the SDRAM controller FIFO. When the tREFI timer expires, a refresh is applied to the SDRAM immediately. Thus, a precise issuing of the refresh in tREFI intervals is achieved at a tiny cost of SDRAM bandwidth.

|  | Avg OnChipProcTime | | Budget | | Total Accesses | | Static Priority |
| --- | --- | --- | --- | --- | --- | --- | --- |
|  | Eq. Density | Incr. Density | Eq. Density | Incr. Density | Eq. Density | Incr. Density | **Only PBS** |
| master1 | $2^3$ | $2^0 \times 2^3$ | $2^2$ | $2^5$ | 2048 | $2^5 \times 100$ | 6 |
| master2 | $2^3$ | $2^1 \times 2^3$ | $2^2$ | $2^4$ | 2048 | $2^4 \times 100$ | 5 |
| master3 | $2^3$ | $2^2 \times 2^3$ | $2^2$ | $2^3$ | 2048 | $2^3 \times 100$ | 4 |
| master4 | $2^3$ | $2^3 \times 2^3$ | $2^2$ | $2^2$ | 2048 | $2^2 \times 100$ | 3 |
| master5 | $2^3$ | $2^4 \times 2^3$ | $2^2$ | $2^1$ | 2048 | $2^1 \times 100$ | 2 |
| master6 | $2^3$ | $2^5 \times 2^3$ | $2^2$ | $2^0$ | 2048 | $2^0 \times 100$ | 1 |

Table 1: Applied Traffic Pattern

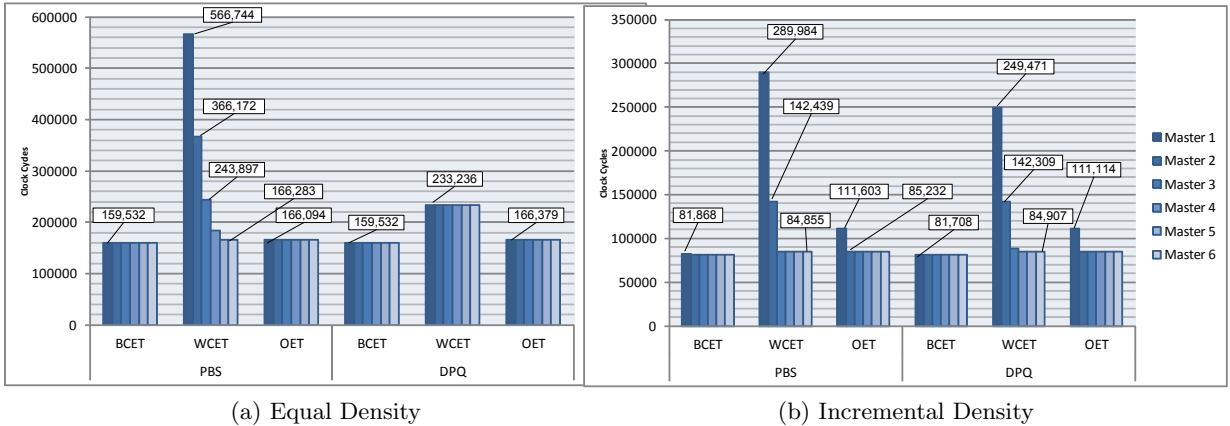

(a) Equal Density  (b) Incremental Density

Figure 7: Results

Although, it is in our future work, we did not integrate our API in WCET analysis tool for this report. Hence, to emulate the behavior of application execution (Fig. 1a) we generated synthetic traces considering them to be the given worst case path. We generated two types of SDRAM traffic and collected traces of them. (1) Equal Density Traffic, where each master generates 2K random alternating accesses to the SDRAM. Average traffic density is identical for all masters. (2) Incremental Density Traffic, where each master generates alternating random accesses such that the traffic density is proportional to its budget. Moreover, the total number of accesses also increases proportionally with traffic density (cf. Table 1). The former traffic pattern is more suitable to the DPQ while the latter is more suitable to the PBS. These tests cover a vast portion of test cases where applications executing on a multi-core architecture have identical bandwidth requirements or there is a huge difference among there bandwidth requirements.

In PBS, master6 has the highest and master1 has the lowest static priority. Although, each master generates alternating read/write accesses, due to the random `OnChipProcTime` worst case scenario during execution is not guaranteed. We calculated how much time will be required by each master to complete its transfers in the worst case.

## 7 Results

For these tests, WCET, BCET and Observed Execution Time (OET) are depicted in Fig. 7a and Fig. 7b. Due to the nature of PBS which prefers a high priority master over lower priority masters, WCET bounds of lower priority masters increase exponentially in Fig. 7a. The WCET bounds produced for PBS vary largely in the equal density test due to the unfairness to lower priority masters while for DPQ, WCET bounds produced for all masters are similar. In the analysis of the DPQ, we assume that the master under investigation is always at the end of the queue when it issues a request. Hence, each master has to wait for five other masters to complete their requests before its own request is scheduled. Such assumption produces similar WCET estimation for all the masters as seen in Fig. 7a. In PBS, highest priority master (master6) has to wait for only one on going low priority transfer. Master5 and master4 have to wait for one low priority plus one high priority, and one low priority plus two high priority transfers respectively. Hence, the WCET bounds produced for high priority masters (master6, master5 and master4) are less in the PBS than in the DPQ for the equal density traffic.

In PBS, for the incremental density traffic (Fig. 7b), the WCET bound produced for master6 is similar to the WCET bound produced by the DPQ although it has the highest priority in PBS. This counter intuitive observation results from the fact that budgets assigned to masters are not identical in this test. Although accesses from master6 are served with low latency, it quickly runs out of the budget and has to wait till the next replenishment period starts. The large WCET bounds of master1 and master2 in both the arbitration schemes are due to the definition of the replenishment period (Equation (1)) which does not include `OnChipProcTime`. Since both of these masters do significantly more accesses in a replenishment period than others, for their late accesses in a replenishment period the condition $(CrtPos + OnChipProcTime) \geq Rp$ of Algorithm 2 is often true. These late accesses are then considered as early accesses in the next replenishment period where they are predicted to experience high interference (cf. Sec. 4.5). In PBS, master1 is further penalized by its lowest priority. The worst case overestimation $WCET/WCET^{nr} \approx 1.04$ for all the masters in both the arbitration schemes.

In the DPQ, for budget accounting, a counter and a comparator are required per master. The queue is constructed from registers and a mux. A counter, a comparator and a register in the queue are added for every additional master which leads to linear increase in the on-chip area. On a Cyclone III FPGA, PBS needs 1551 logic elements(LE) while DPQ needs 1746 LEs for 6 masters at 125 MHz. Although the DPQ arbiter is slightly larger than the PBS, since an arbiter is a tiny component of the system, the increase in area of the entire system is negligible.

## 8 Conclusion

This report introduces a novel arbitration scheme called Dynamic Priority Queue - DPQ for shared SDRAM in an MPSoC. The DPQ enables low WCET bounds and satisfies distinct bandwidth requirements of masters. An algorithm to calculate worst case access latency is presented and implemented as an API model for integration into third party WCET analyzers or virtual platforms. The algorithm calculates the WCET in isolation assuming no knowledge of co-existing applications. WCET of benchmarks was calculated using the proposed API and verified on a commercial FPGA board. The results show that the DPQ produces lower WCET bounds than the Priority Based Budget Scheduler - PBS even in a test which is biased towards the PBS arbitration scheme.